\documentclass[lettersize,journal]{IEEEtran}
\usepackage{amsmath,amsfonts}
\usepackage{algorithmic}
\usepackage{algorithm}
\usepackage{array}
\usepackage{hyperref}
\usepackage{makecell}
\usepackage{graphicx}
\usepackage{textcomp}
 
\usepackage{longtable}
\usepackage{booktabs}
\usepackage{float}

\usepackage{longtable}
\usepackage[caption=false,font=normalsize,labelfont=sf,textfont=sf]{subfig}
\usepackage{textcomp}
\usepackage{stfloats}
\usepackage{url}
\usepackage{verbatim}
\usepackage{graphicx}
\usepackage{cite}
\hypersetup{
colorlinks=true,
linkcolor=black,
citecolor=black
}
\begin{document}

\title{Ultra-Wideband Technology: Characteristics, Applications and Challenges}

\author{Chutao Zheng\\~\IEEEmembership{School of Integrated Circuit Science and Engineering 
(Exemplary School of Microelectronics)\\ University of Electronic Science and Technology of China}\\ \hspace*{\fill} \\
Yuchu Ge\\~\IEEEmembership{School of Integrated Circuit Science and Engineering 
(Exemplary School of Microelectronics)\\ University of Electronic Science and Technology of China}\\\hspace*{\fill}\\
Anfu Guo\\~\IEEEmembership{School of Electronic Science 
and Engineering\\University of Electronic Science and Technology of China}
}

\maketitle

\begin{abstract}
Ultra-wideband (UWB) technology is a wireless communication technology designed for short-range applications. It is characterized by its ability to generate and transmit radio-frequency energy over an extensive frequency range. This paper provides an overview of UWB technology including its definition, two representative schemes and some key characteristics distinguished from other types of communication. Besides, this paper also analyses some widely used applications of UWB technology and highlights some of the challenges associated with implementing UWB in real-world scenarios. Furthermore, this paper expands upon UWB technology to encompass terahertz technology, providing an overview of the current status of terahertz communication, and conducting an analysis of the advantages, challenges, and certain corresponding solutions pertaining to ultra-wideband THz communication.
\end{abstract}


\section{Introduction}
Apple's release of the iPhone 11 with an ultra-wideband (UWB) transceiver in 2019 has caused the ever-unknown ultra-wideband technology to receive widespread attention. This technology provides users with the ability to locate their position more accurately, allowing for much improved and expanded applications in areas such as indoor navigation, indoor positioning, and Internet of Things (IoT) connectivity. UWB is a wireless communication technology characterized by transmitting data by transmitting a continuous pulse signal in a very short period of time\cite{2009}. Initially, Ultra-Wideband technology was primarily used in radar and military communications to play an important role in stealth operations and precision ranging\cite{Tutorial}. By 2002, however, UWB technology was moving into the civilian world. In that year, the U.S. Federal Communications Commission (FCC) allowed the use of UWB technology in the broadband range, which promoted the development of UWB technology in the field of communications. Due to its broadband nature, UWB technology can offer higher data transfer rates and faster response times, making it one of the ideal choices for future mobile devices.

With an extremely wide bandwidth and low power spectral density, UWB communication can realize high data transmission rate, high confidentiality communication, and accurate ranging and positioning ability, meanwhile, it also has low interference to other frequency bands. However, because of factors such as high price and limited distance, it is difficult to be very widely used, thus its development is limited. Nowadays, as an enabler for supporting Internet of Things (IoT) networks and applications\cite{yin1}, the fifth-generation (5G) wireless network has been developed rapidly, thus contributing to the development of the IoT. And the emergence of new IoT services and applications, such as remote robotic surgery and flying vehicles, also requires higher transmission rate to improve the quality of IoT service delivery and business \cite{yin2}. However, due to the limitations of the bandwidth, Wi-Fi and Bluetooth will not be able to meet the transmission rate requirements. On the contrary, UWB communication has a very wide bandwidth to meet the rate requirement and will likely receive more attention. There are also many other features such as low power consumption, low interference, high interference immunity and high security. The detail of these characteristics and corresponding principles will be discussed in Section\ref{section5} and Section\ref{section6}. In addition, with high confidentiality and precise accurate and positioning ability, the use of UWB communication will also enhance the confidentiality and accuracy of the IoT. Therefore, with the continuous development of IoT technology, UWB communication will also be more widely used and rapidly developed.

Section\ref{section2} provides an overview of UWB technology, including its definition, frequency range, and maximum radiated power, and compares UWB with narrowband and broadband from the frequency domain and bandwidth perspectives. Then Section\ref{section3} introduces IR-UWB, a representative of the implementation scheme of UWB technology. This section firstly provides a general introduction of the IR-UWB concept, then compares the conventional signaling techniques with broadband signaling techniques, and finally describes the steps of the IR-UWB scheme, including a basic overview of the key steps such as signal modulation, pulse expansion, and so on.

Section \ref{section4} describes another representative of the UWB technology implementation scheme, MB-OFDM-UWB. This section begins with an overall overview of the IR-UWB scheme. This section introduces the concept and vision of MB-OFDM-UWB in general, then briefly describes the implementation methodology of MB-OFDM-UWB, and finally describes its features, including avoidance of used frequencies, strong multipath resistance, capture capability and strong spectrum flexibility and coexistence. 
Section \ref{section7} gives some widely used ranging and positioning methods. For ranging, Two-Way Ranging (TWR) and Double-side Two-Way Ranging (DS-TWR) are discussed to solve the error caused by clock drift and time asynchronization.  This section focus on the principle and advantages and disadvantages of several basic geometry-based methods are introduced. In addition, some advanced methods that can be used in NLOS environment are mentioned. Finally, Section \ref{section8} gives a brief overview of the current state of the UWB communication and THz band, then compares UWB-THz communication with other communication methods, analyzes its advantages, and introduces some challenges faced in long-distance communication and corresponding solutions.

\section{Overview of UWB}
\label{section2} 
\subsection{Definition of UWB}
According to ITU-R SM.1755-0 recommendation, UWB communication technology is defined as follows: UWB technology is a wireless communication technology specifically designed for short-range applications. It encompasses the intentional generation and transmission of radio-frequency energy, which propagates over an extensive frequency range and is potentially coexistent with multiple frequency bands allocated for radiocommunication services. UWB-enabled devices commonly feature antenna radiation intentionally designed to exhibit a minimum –10 dB bandwidth of 500 MHz or a –10 dB fractional bandwidth exceeding 0.2\cite{ITU}.

%\subsection{Frequency Range}
According to FCC rules, UWB operates in a frequency ranging from 3.1 GHz to 10.6 GHz, with a bandwidth of 7.5 GHz\cite{FCC}.

\subsection{Maximum Radiated Power}
\begin{figure}[!ht]
        \centering
        \includegraphics[width=0.99\linewidth]{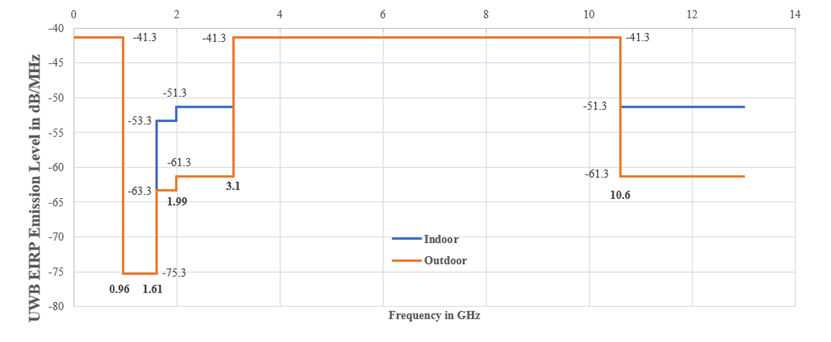}
        \caption{FCC emission limits\cite{FCC}.}
        \label{z1}
\end{figure}

To protect other radio services from UWB interference, the FCC has assigned conservative emission masks for UWB devices. Within the frequency range of 3.1 GHz to 10.6 GHz, the provisions outlined by these regulations enforce constraints on the maximum radiated power exhibited by UWB signals under various operational scenarios. Fig.\ref{z1} illustrates the emission limits imposed by these regulations on the UWB devices employing indoor and outdoor implementations\cite{FCC}.

The template that restricts the spectrum occupancy of UWB communication systems is commonly referred to as the UWB spectrum mask. Its formulation in different countries ensures that UWB systems do not cause strong interference to neighboring frequency bands when utilizing the spectrum.

\begin{figure}[!ht]
        \centering
        \includegraphics[width=0.85\linewidth]{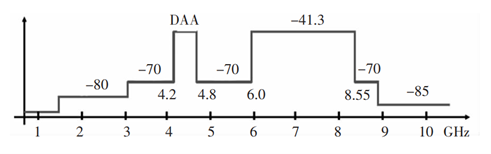}
        \caption{UWB spectrum mask in China\cite{2009}.}
        \label{z2}
\end{figure}

Fig.\ref{z2} presents the UWB spectrum mask adopted in China. In the UWB spectrum mask, the horizontal axis represents frequency, typically measured in gigahertz (GHz), while the vertical axis represents power, commonly measured in decibel milliwatts (dBm). In Fig.\ref{z2}, the  “DAA" stands for “Detection and Avoidance", which represents a requirement within the UWB spectrum mask. It indicates that UWB devices must possess the capability to detect and avoid interfering with other wireless systems when utilizing the spectrum.

\subsection{Comparison with Narrowband and Broadband}
% \label{section3} 
In terms of the frequency domain, UWB signals are significantly different from traditional narrowband and broadband signals due to their wider bandwidth. A signal with a relative bandwidth of less than 1$\%$ is deemed narrowband if we use relative bandwidth as the measure. Broadband signals have a relative bandwidth of between 1$\%$ and 20$\%$. UWB signals have a relative bandwidth larger than 20$\%$ or absolute bandwidth higher than 500MHz\cite{guanjian}.

Besides, UWB signals exhibit a distinct characteristic of having a wide frequency spectrum with energy distributed uniformly across the entire frequency range, resulting in a flat spectral profile. In contrast, narrowband signals possess a narrower frequency bandwidth and typically exhibit prominent power peaks concentrated around their center frequency, indicating discrete spectral peaks. While broadband signals generally have a wider frequency bandwidth, they do not match the spectral spread achieved by UWB transmissions. This is because UWB signals can span a frequency spectrum width exceeding 500 megahertz (MHz), extending over several GHz. Consequently, UWB signals surpass the conventional notion of broadband transmissions in terms of spectral width.

\begin{figure}[!ht]
        \centering
        \includegraphics[width=0.9\linewidth]{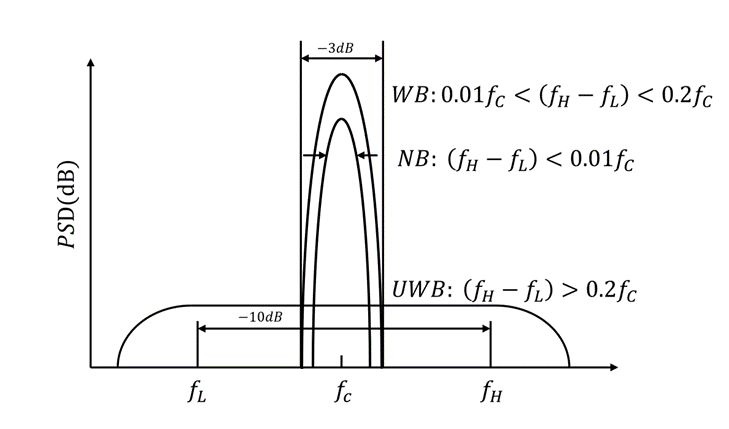}
        \caption{Comparation among three communication forms.\cite{tang}}
        \label{z3}
        \end{figure}

The Fig.\ref{z3} provides a rough illustration of the relative bandwidth and power spectral density for the three communication forms mentioned above.

\section{Impulse Radio UWB (IR-UWB)}
\label{section3} 
\subsection{IR-UWB Technology}
Impulse radio refers to the wireless technology that utilizes impulse pulses (ultra-short pulses) as information carriers. The modulation of extremely thin pulses (often less than 1 ns) to get wide bandwidth data for transmission is the primary aspect of this technique. IR-UWB is a representative implementation scheme in the field of UWB\cite{tang}.

\subsection{Comparison between IR-UWB Signals and Traditional Communication Signals}
\begin{figure}[!ht]
        \centering
        \includegraphics[width=0.8\linewidth]{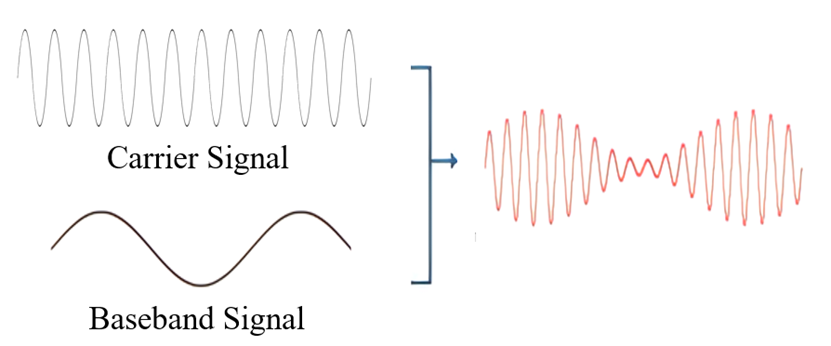}
        \caption{An amplitude-modulated signal.}
        \label{z4}
        \end{figure}

Traditional communication systems use continuous wave transmissions, which are generated by a local oscillator and consist of a continuous high-frequency carrier wave. Information is transmitted by modulating the carrier wave with techniques such as amplitude modulation (AM) and frequency modulation (FM) and then delivering it via an antenna. This technique of wireless communication is used by current wireless broadcasting, 4G communication, and Wi-Fi. Fig.\ref{z4} depicts the production of a continuous wave signal for use in amplitude modulated voice transmission.

\begin{figure}[!t]
        \centering
        \includegraphics[width=0.8\linewidth]{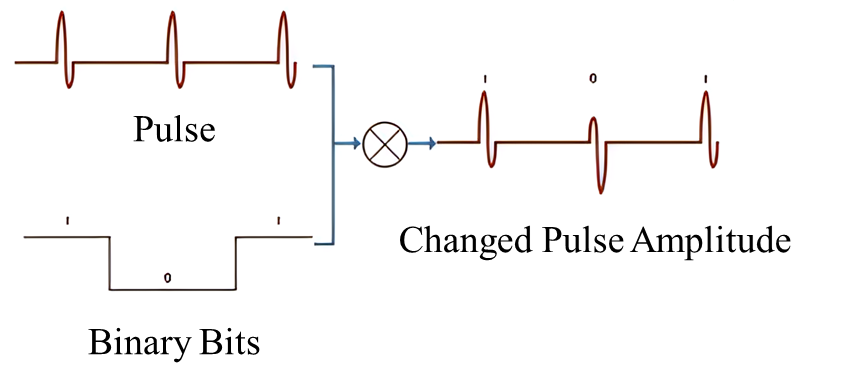}
        \caption{A phase-modulated IR-UWB signal.}
        \label{z5}
        \end{figure}

IR-UWB transmissions, on the other hand, do not necessitate the production of continuous high-frequency carrier waves. At the most basic level, they just need to generate pulses with durations as short as nanoseconds, which will subsequently be sent via an antenna. Information can be loaded by varying the amplitude, length, and phase of the pulses, resulting in information transmission. The Fig.\ref{z5} depicts the generation of an IR-UWB signal for conveying binary zero code via phase modulation\cite{abrief}.

\subsection{Implementation Approach of IR-UWB}
The following is the Implementation Approach of IR-UWB's steps.
\subsubsection{Selection of pulse shape}
 IR-UWB signals employ pulse signals as the fundamental modulation waveform. Commonly used pulse shapes include Gaussian Pulses, Binomial Pulses, etc. To effectively radiate the signal energy through the antenna, certain requirements are placed on the frequency spectral characteristics of the chosen pulse waveform (i.e., absence of DC component, minimal presence of low-frequency components, with predominant signal energy concentrated in the RF region). Therefore, pulse radio utilizes derivatives of Gaussian functions as the transmitted pulse waveform, allowing for different bandwidths and center frequencies to be achieved by selecting the pulse width and sequence.

\subsubsection{Generation of pulse sequence}
The pulse sequence essentially represents the discrete form of the IR-UWB signal. Hence, it is necessary to generate the pulse sequence based on the digital information to be transmitted and associate each pulse with its corresponding binary bit in the subsequent steps.

\subsubsection{Modulation of pulse sequence}
Modulation of the pulse sequence essentially involves loading the digital information into the pulse sequence through the specific modulation method. Information transfer is achieved by directly modifying the amplitude, duration, and phase of the pulses for information loading.

\begin{figure}[!t]
        \centering
        \includegraphics[width=0.8\linewidth]{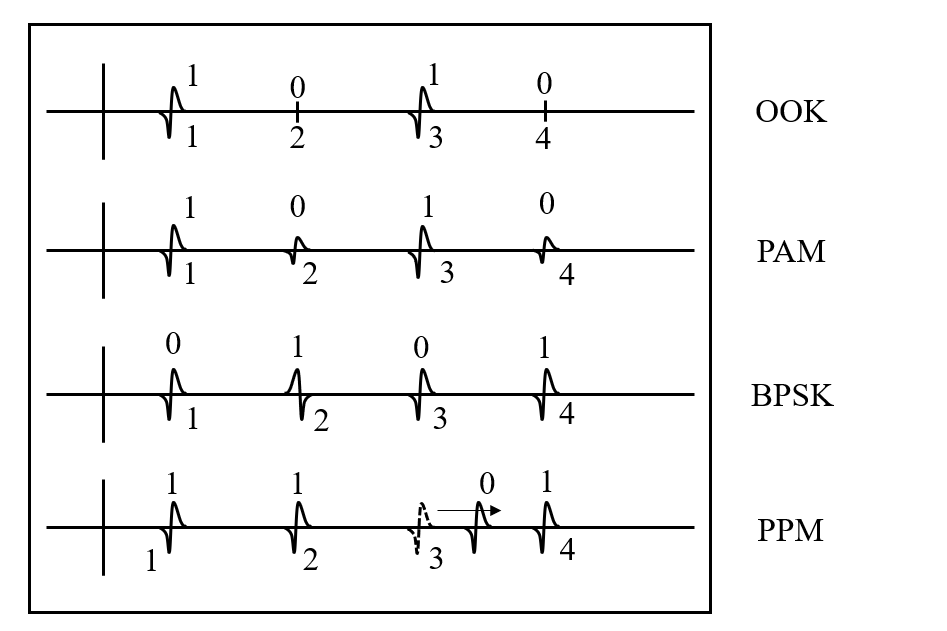}
        \caption{Four common modulation methods for IR-UWB signals\cite{abrief}.}
        \label{z6}
        \end{figure}

Then four common modulation methods will be presented: Pulse Position Modulation (PPM), Pulse Amplitude Modulation (PAM), On-Off Keying (OOK) and Binary Phase Shift Keying (BPSK) \cite{abrief}. PPM represents digital information by adjusting the position of the pulse, while PAM represents digital information by adjusting the amplitude of the pulse. On-Off Keying (OOK) represents digital information by adjusting the presence or absence of the pulse, whereas Binary Phase Shift Keying (BPSK) represents digital information by adjusting the phase of the pulse. The Fig.\ref{z6} provides a rough illustration of these modulation techniques\cite{abrief}.

\subsubsection{Amplitude adjustment}
Amplitude adjustment refers to the process of adjusting the amplitude of the modulated pulse sequence to meet the transmission requirements. By modifying the amplitude of the pulse, the energy level of the signal can be changed, thereby enhancing the transmission distance of the signal. Increasing the amplitude of the pulse can provide stronger energy transmission, enabling the signal to be received at greater distances. Therefore, controlling the transmission power is crucial in certain applications such as wireless sensor networks or energy-constrained devices. In addition, by adjusting the amplitude of the pulse, the power level of the signal can be controlled according to system requirements. This helps reduce energy consumption, prolong battery life, and meet specific transmission needs.

\subsubsection{Pulse extension}
Pulse extension refers to the process of extending the pulse width of the signal from the original narrow pulse to a wider pulse signal. Common pulse extension techniques include chip extension and filtering extension. Chip extension involves performing mathematical operations between the original narrow pulse and a longer pseudo-random sequence (known as a chip sequence) to generate a wider expanded pulse. As a result, the expanded pulse has a broader bandwidth in the frequency domain, making it easier for the receiver to distinguish the signal and enhancing the signal's immunity to interference. Another technique is filtering extension, which utilizes filtering operations to widen the pulse signal. 

Pulse extension is a commonly used technique in IR-UWB systems to increase the temporal duration and energy of the pulse signal, thereby improving the transmission distance and resistance to multipath interference. It helps mitigate the effects of multipath interference. In terms of long-distance transmission, pulse extension offers better transmission distances. The increased energy and improved interference resistance of the extended pulse enable the signal to be transmitted over longer distances, mitigating the effects of signal attenuation.

\subsubsection{Transmission of pulse signals}
The modulated pulse sequence is transmitted through the antenna, forming a wireless signal. This signal propagates through the air and can be captured by the receiver.

\subsubsection{Signal reception}
The receiver antenna of the IR-UWB system receives the transmitted pulse signal from the transmitter. The received RF signal is amplified by the receiver for subsequent processing. In an IR-UWB system, the received signal is often weak, and therefore, it needs to be amplified to an appropriate signal level by the RF front-end amplifier for subsequent demodulation and processing.

\subsubsection{Signal demodulation}
After amplification by the RF front-end, the signal needs to be demodulated and filtered to extract the pulse signal. Common demodulation techniques for Differential Binary Phase Shift Keying (DBPSK) include Energy Detection Based Demodulation (EDD) Technique, Correlator Detection Based Demodulation (CDD) Technique and Windowing Detection Based Demodulation (WDD) Technique\cite {noncdemoDBPSK}. Common demodulation schemes for PPM include Energy Detection Scheme, Energy Detection Scheme, Correlator Scheme and Windowing Scheme\cite{noncdemoPPM}.

\subsubsection{Filtering}
After demodulation, the resulting baseband signal may contain noise. Filtering operations are performed to remove these unwanted components. Different types of filters can be selected, such as low-pass filters to remove high-frequency noise or band-pass filters to select signals within specific frequency ranges\cite{MBlater}. 

\subsubsection{Reconstruction of short pulse sequences}
Based on the demodulated signal, the original short pulse sequence can be reconstructed using signal processing algorithms. This may involve using threshold detectors or energy detectors to extract the pulses and performing further processing and recognition as required.

\section{MB-OFDM-UWB}
\label{section4}
\subsection{MB-OFDM-UWB Technology}
With the development of UWB technology, new UWB techniques have emerged, among which multi-band technology is the most representative. Multiband Orthogonal Frequency Division Multiplexing UWB (MB-OFDM-UWB) is a prime example that combines this technique. In an MB-OFDM-UWB system, the available UWB spectrum is divided into several subbands, each with a width no less than 500 MHz. During communication, a portion or all of the subbands can be dynamically utilized based on the information rate, system power consumption requirements, coexistence with other systems, etc., to improve the spectrum utilization by simultaneously transmitting UWB signals in different frequency bands. These UWB signals do not interfere with each other because they have different frequencies \cite{MBlater}.

Due to its significant similarities with traditional technologies, this paper mainly introduces the general scheme and characteristics of the MB-OFDM-UWB technology, without delving into specific implementation steps.

\subsection{MB-OFDM-UWB Scheme}
\begin{figure*}[!t]
        \centering
        \includegraphics[width=0.7\linewidth]{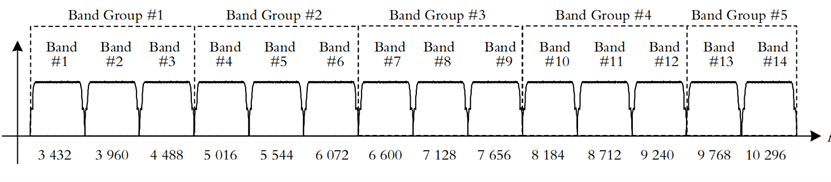}
        \caption{MB-OFDM-UWB scheme\cite{2009}.}
        \label{z7}
        \end{figure*}

The MB-OFDM-UWB scheme adopts a multi-band approach, as illustrated in Fig.\ref{z7}. The 3.1 GHz to 10.6 GHz frequency range is divided into 14 subbands (Bands), with each subband of 528 MHz used to transmit OFDM signals consisting of 128 subcarriers. Each subcarrier occupies a bandwidth of 4.125 MHz. The 14 subbands are further divided into 5 subband groups (Band Groups), with each group containing three or two subbands, enabling frequency diversity through hopping between different subbands\cite{2009}.

The MB-OFDM-UWB scheme possesses three main characteristics:

\begin{itemize}
\item Avoidance of Used Frequencies:

MB-OFDM-UWB divides the spectrum into multiple subbands and utilizes orthogonal frequency division multiplexing (OFDM) on each subband. As different subbands are orthogonal to each other, they can simultaneously transmit independent data streams without mutual interference within the same spectrum. This design enables MB-OFDM-UWB to be flexible in spectrum selection, allowing it to avoid frequencies already in use by other wireless technologies and reducing interference with other systems\cite{tang}.

\item Strong Multipath Resistance and Capture Capability:

The multi-band and orthogonal frequency division multiplexing techniques employed in MB-OFDM-UWB allow for the utilization of multipath propagation during transmission. As the different subbands are orthogonal, the different signals formed by multipath propagation within them cancel each other out, reducing the impact of multipath interference on the system. Moreover, MB-OFDM-UWB exhibits good resolution in the time domain, enabling it to better distinguish different paths in multipath propagation. This enhances the system's capability to capture and resolve multipath signals, improving resistance to multipath interference.

\item Strong Spectrum Flexibility and Coexistence:

The multi-band and orthogonal frequency division multiplexing techniques in MB-OFDM-UWB provide strong spectrum flexibility. Different subbands can be flexibly configured according to various application requirements and environmental conditions. Furthermore, due to the orthogonality between subbands, multiple MB-OFDM-UWB devices can coexist in relatively close proximity without interfering with each other. This results in good coexistence capabilities in multi-user scenarios, making MB-OFDM-UWB suitable for high-density wireless communication environments\cite{2009}.
\end{itemize}

However, the MB-OFDM-UWB scheme does have its limitations. It utilizes conventional OFDM technology instead of the pulse-based technology, which leads to higher power consumption compared to IR-UWB. Additionally, it lacks the penetrating capabilities and high security features of IR-UWB\cite{2009}.

\section{Key Characteristics of UWB Technology}
\label{section5} 
In this section, some features of UWB will be further discussed. These characteristics make UWB technology have strong prospects for applications in certain specific fields.

\subsubsection{Low power consumption}

Pulse-based UWB technology often utilizes intermittent pulse transmission for data communication. In UWB positioning systems, the operating cycles are typically in the range of 0.2 ns to 1.5 ns, resulting in low power consumption. In high-speed communications, UWB devices consume only tens to hundreds of watts \cite{chai,zhao2022wknn}. Therefore, compared to conventional wireless devices, UWB positioning devices offer better battery life and lower electromagnetic radiation.

\subsubsection{Minimal interference with other devices}

UWB technology shares spectrum resources with other wireless communication systems. It utilizes a spectrum ranging from 3.1 GHz to 10.6 GHz, with a bandwidth of up to 7.5 GHz, without requiring dedicated or proprietary frequency bands. By limiting the transmission power, UWB also avoids causing interference with other systems. This flexible spectrum utilization is one of the main reasons for the significant development of UWB technology, especially in situations where spectrum resources are scarce\cite{guanjian, g16}.

\subsubsection{Strong interference (from other devices) resistance}

Due to the wide frequency spectrum distribution of UWB signals, which occupy a large frequency range, UWB signals exhibit unique characteristics in the frequency domain and possess strong resistance to narrowband interference signals. The signal energy is spread across the frequency spectrum, making it more resilient against interference\cite{chai}.

\subsubsection{High data rates}

\begin{equation}
        C=B log_{2}{(1+\frac{S}{N})}
\end{equation}
As the Shannon Formula shows above, the maximum achievable data rate or channel capacity ($C$) of a communication channel can be calculated using Shannon's channel capacity formula, where $B$ represents the channel bandwidth, $N$ represents the Gaussian white noise power spectral density, and $S$ represents the average signal power. By increasing the transmission bandwidth, UWB technology achieves extremely high data rates. Typically, the maximum data transmission speed can reach from several tens of megabits per second to several hundreds of megabits per second.
 
\subsubsection{High security}

Compared to non-UWB radio communication signals, UWB signals may be more discreet and harder to detect\cite{guanjian}. This is because UWB signals occupy a wide frequency band and can resemble noise-like signals. In other words, within such a wide frequency range, UWB signals are masked by ambient noise, making them difficult to detect. Each bit is typically represented by a large number of very low amplitude pulses (often below the noise level). These characteristics enable secure transmission with low detection probability (LPD) and low intercept probability (LPI).

\subsubsection{Time resolution and robustness against multipath interference}

In traditional wireless communication systems, most RF signals are continuous waveforms, resulting in the effects of multipath propagation on signal quality and transmission rate. With UWB wireless transmission, short-duration periodic pulses are used, typically with pulse widths ranging from a few nanoseconds to a few tens of picoseconds. These pulses have small duty cycles and low work periods, preventing overlap of multipath signals in time domain. However, due to the extremely short pulse width, UWB signals appear as highly sharp pulses in the time domain. The signals reflected from surfaces between the transmitter and receiver are unlikely to overlap in time\cite{ref10}. By using time-domain processing techniques, UWB receivers can distinguish between direct and reflected signals, effectively mitigating the adverse effects of multipath interference\cite{chai}.

\subsubsection{ Low signal attenuation and strong penetration}

IR-UWB signals cover a wide frequency spectrum. This implies that IR-UWB signals simultaneously utilize a broad range of frequency resources during transmission. Compared to traditional narrowband signals, the energy of IR-UWB signals is more dispersed across the spectrum, resulting in relatively lower signal power at individual frequencies and reducing the impact of frequency selective fading\cite{ITU}.

\begin{table*}%星号表示双栏
\renewcommand\arraystretch{1.5}%保证每列高度是原先的1.5倍
\caption{Comparison of Different Wireless Communication Technologies\cite{duibi}}
 \begin{tabular}{p{1.2cm}<{\raggedright}p{1.5cm}<{\raggedright}p{1cm}<{\raggedright}p{1.8cm}<{\raggedright}p{1.5cm}<{\raggedright}p{1.5cm}<{\raggedright}p{1.5cm}<{\raggedright}p{4.5cm}<{\raggedright}}%
 
  \Xhline{1.2pt}
       \textbf{Technology}
       & \textbf{Data Rate} & \textbf{Range} & \textbf{Accuracy} & \textbf{Network Topology}  & \textbf{Complexity   } &\textbf{Power Consumption }&\textbf{Typical Applications} \\ 
  \Xhline{1.2pt}%

  \textbf{Bluetooth }& 1 Mbits/s & 10 meters & Several meters& Ad-hoc; very small networks   & High& Medium & Wireless connectivity between devices such as phones, PDA, laptops, headsets \\ \hline
   \textbf{Wi-Fi} &11 \& 54 Mbits/s& 50-100 meters & Several meters & Point to hub  & High & High & Wireless LAN connectivity; broadband Internet access\\ \hline
    \textbf{ZigBee}& 20, 40 and 250 Kbits/s  & 10-100 meters&0.5 meters& Ad-hoc; peer to peer star; mesh & Low& Very low  & Industrial control and monitoring; sensor networks; building automation; home control; automation, toys, games     \\ \hline
     \textbf{UWB}&50-100Mb/s, $>$500Mb/s expected in future  &170 meters&5 to 10 cm&Point to point&Medium&Low&Industrial control and monitoring; sensor networks; building automation; home control streaming video; home entertainment applications\\
  \Xhline{1.2pt}%第三条粗线
  \end{tabular}

\end{table*}

\section{Comparison with Other Short-range Wireless Technologies}
\label{section6} 
The advantages of UWB can be observed by comparing it with commonly used short-range wireless technologies. The specific comparisons are presented in the Table.\uppercase\expandafter{\romannumeral1}. The advantages of UWB in terms of accuracy are evident. UWB technology exhibits a remarkable capability to measure distance and location with an impressive accuracy range of 5 to 10 cm. In contrast,  wireless systems such as Wi-Fi, Bluetooth, and others can only achieve accuracy at the level of several meters.

Furthermore, UWB stands out with significantly lower power consumption compared to Wi-Fi. However, it is worth noting that one tradeoff of UWB is its limited compatibility and interaction capabilities with contemporary smartphones and tablets, where Wi-Fi and Bluetooth-enabled devices excel. Nevertheless, certain companies are proactively adopting UWB by developing hybrid devices that integrate both UWB and either Wi-Fi or Bluetooth technologies, aiming to combine the best features and advantages of each technology.

\section{Applications of UWB Technology}
\label{section7}
UWB technology is considered as one of the most promising radio technologies due to its ultra-wide signal bandwidth, low transmit power consumption and high data rate. 
Currently, IR-UWB technology is widely used in applications related to precise ranging, positioning, and data transferring. 

Indoor scenarios make UWB technology more competitive in positioning and ranging. Compared to outdoor environments, indoor settings are more complex due to the presence of numerous objects that cause signal reflections, resulting in multipath and delay issues. Additionally, the indoor environment often leads to non-line-of-sight (NLoS) propagation, where some signals cannot directly propagate from the transmitter to the receiver in a line-of-sight path, leading to inconsistent time delays at the receiver. UWB transmission involves very short pulses, providing an advantage in the resolvability of multipath components and can partially alleviate multipath interference.

In this section, some methods addressed to ranging and positioning are introduced, and the advantages and weaknesses of UWB technology compared with other technologies are analyzed.  

\subsection{UWB Ranging Techniques}

\begin{figure*}[!t]
        \centering
        \subfloat[]{\includegraphics[width=0.4\linewidth]{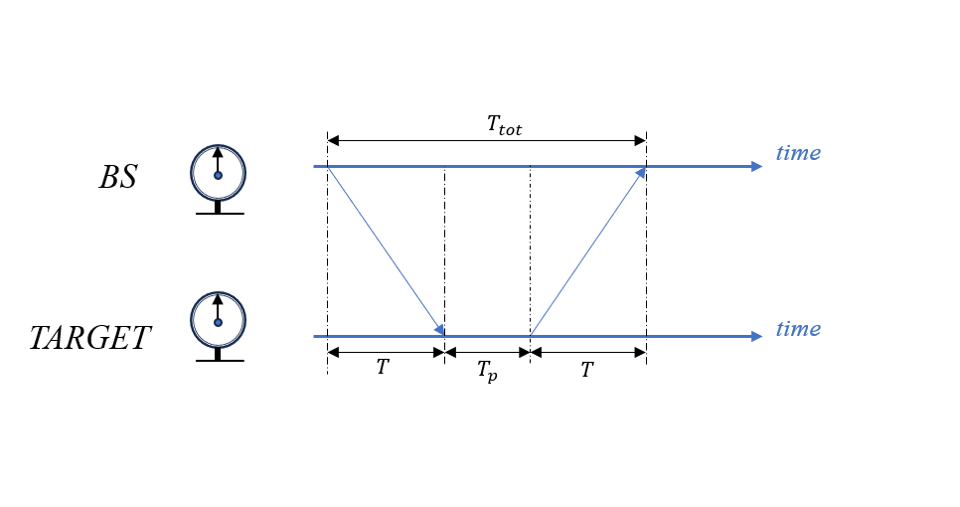}%
        \label{11}
        }
        \hfil
        \subfloat[]{\includegraphics[width=0.4\linewidth]{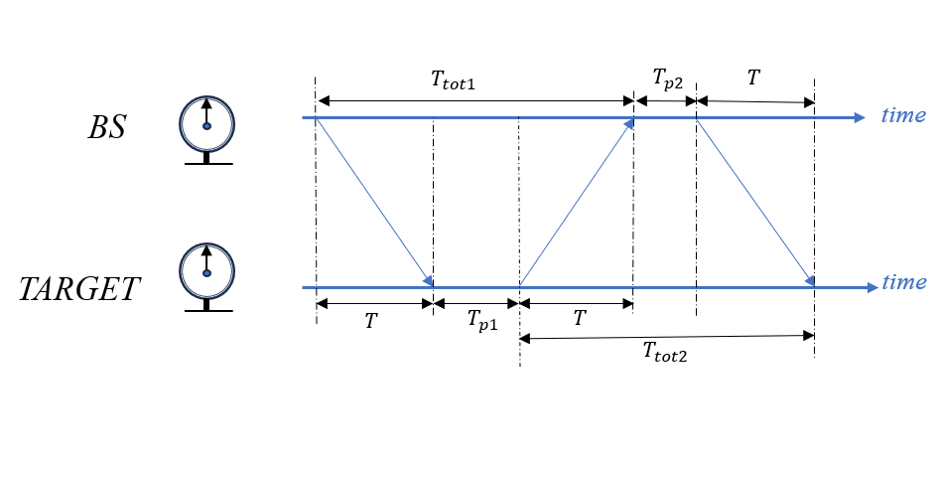}%
        \label{22}
        }
        \caption{Two types of two way ranging.}
        \label{fig_sim}
\end{figure*}
In distance measurement, the “Two Way Ranging" algorithm is proposed to calculate the time of flight ($T_f$). By combining the signal's velocity in the environment, the distance between the target and the base station can be accurately determined. Assuming that the device in tag processes information for a time period $T_p$ and the base station experiences a time period $T_{tot}$ from transmitting the signal to receiving the response signal,
the round-trip time between the localization tag and the base station can be represented as 
 \begin{equation}
       \hat{T_f}  =\frac{(T_{tot} - T_p)}{2} .
 \end{equation}     

The distance between the tag and the base station, 
denoted as D, is calculated as $D = v\times   T_f$, where $v$ represents the speed of electromagnetic wave propagation in the air. 
 Getting a TOF with high precision requires strict synchronization between the transmitter and receiver (located within the base station) 
 and often necessitates the transmission of timestamps along with the signals (depending on the underlying communication protocol)\cite{ref6}.

 In the process of measuring the Time of Flight ranging using this method, the biggest advantage is that both time differences ($T_{tot},T_{P}$) are calculated based on the local clock. 
 That means, the local clock errors can be compensated for, but there may be slight clock frequency drift between different devices.
 Assuming that the frequency drift for the base station and the target are denoted as $e_A$ and $e_B$, respectively. Consequently,
 the measuring error of $T$ will increase with the increase of $T_p$. The equation for the ranging error can be expressed as follows\cite{ref11}:
 \begin{equation}
        e =\hat{T_f}-T_f=\frac{(e_A-e_B)\times T_p}{2}+e_A\times T_f.
 \end{equation}

 Furthermore, some additional improvements can be made to the computation of $T_f$ to further reduce errors. 
 For instance, the previously mentioned approach can be modified by adopting a method called Double-sided Two-way Ranging (DS-TWR). 
 This method mitigates the effect of clock drift by leveraging an additional message transmitted from the anchor to the tag again, as depicted in Fig.\ref{22}. 
 By mathematically simplifying the process, we can derive the following computation results:\cite{ref11} 
\begin{equation}
        T_f=\frac{T_{tot1}\times T_{tot2}-T_{p1}\times T_{tot2}}{T_{tot1}+T_{tot2}+T_{p1}+T_{p2}}
 \end{equation} 

 The error is determined by one of the following two factors, depending on which side has a smaller clock drift:\cite{ref11} 
 \begin{equation}
        T_f=e_A \times T_f \qquad
        T_f=e_B \times T_f
 \end{equation} 

 By comparison, both conventional TWR and the proposed alternative method (DS-TWR) aim to reduce errors caused by clock drift. However, the advantage of the latter is that the error only relies on the clock drift of a single device. If one device has a significantly better time reference compared to the other, it will enhance performance\cite{ref11}.

 The key factors influencing TWR accuracy are signal bandwidth and sampling rate. 
 Low sampling rates (in the time domain) decrease the TWR resolution as the signal may arrive between sampling intervals.
 Frequency-domain super-resolution techniques are commonly used to achieve high-resolution TWR from the channel frequency response. 
 In indoor environments with multipath, larger bandwidth leads to higher TWR estimation resolution. Although larger bandwidth and super-resolution techniques can enhance TWR performance, 
 they cannot eliminate significant positioning errors when there is no direct line of sight between the transmitter and receiver. 
 This is because obstacles cause signal deviation, resulting in longer signal paths and increased signal flight time\cite{ref6}. 

Therefore, when addressing non-line-of-sight issues, utilizing machine learning to distinguish 
line-of-sight signals and correcting errors to enhance ranging accuracy can be of some progress. 
For instance, employing machine learning in conjunction with Multiple-Input Multiple-Output (MIMO) systems 
enables optimization of the signal transmission process. Reinforcement learning algorithms can adjust 
antenna directions and power weights to achieve improved transmission performance\cite{ref9}. Besides,
 leveraging machine learning models to analyze and exploit the multipath propagation characteristics
  of signals in non-line-of-sight transmission can further enhance performance. 
  In scenarios where signals may be subject to interference and damage during non-line-of-sight 
  transmission, machine learning algorithms can aid in data recovery and error correction. 
  By learning signal patterns and features, these algorithms aim to restore and repair the damaged data as accurately as possible.

\subsection{UWB-based Positioning Techniques }
When applicating UWB in positioning and ranging, geometric-based methods including Time of Arrivals(TOA) positioning, 
Time Difference of Arrival (TDOA) (positioning, Angle of Arrival (AOA) positioning, and Two-Way Ranging (TWR) positioning are widely used to achieve the goal. 
In practice, people also combine multiple geometric algorithms for localization, enhancing both accuracy and interference resistance\cite{ref9}. However, this comes at the expense of increased equipment costs. Next, the principles of several fundamental algorithms will be introduced\cite{ref7,ref8,ref9}.

\subsubsection{TOA} 

\begin{figure}[!t]
        \centering
        \includegraphics[width=0.85\linewidth]{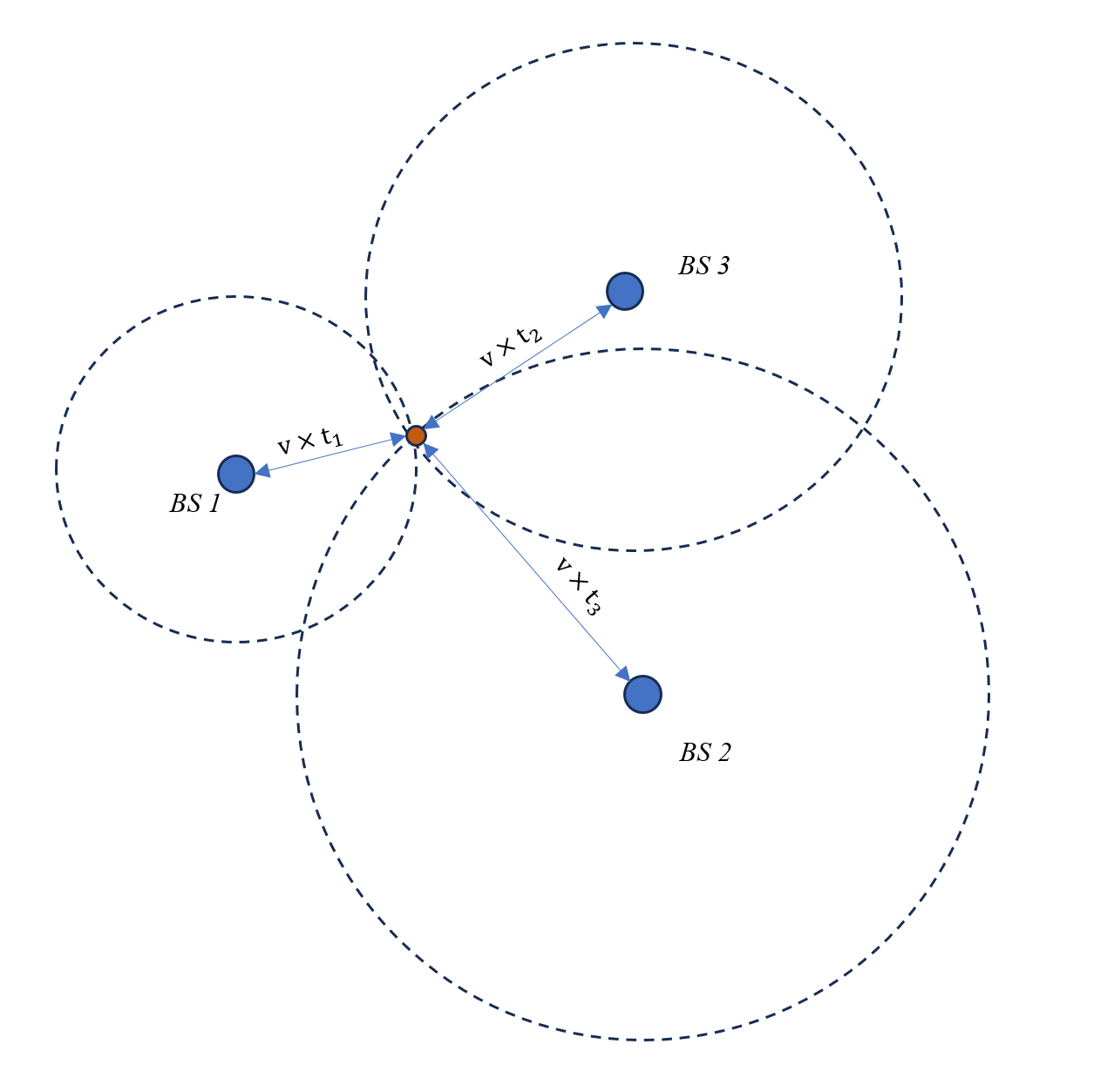}
        \caption{TOA based positioning.}
        \label{fig1}
        \end{figure}

The TOA algorithm utilizes the time information of the signals received by the base stations to perform calculations. For two-dimensional(2-D) positioning, this algorithm requires a minimum of three coordinated base stations for its application. By utilizing three sets of arrival times, we can calculate the distances between the target and the three base stations, enabling determine the target's location, as Fig.\ref{fig1} illustrates. This method necessitates strict synchronization of the clocks between the target and the three base stations because the UWB technology used in this approach employs narrow pulse signals that can achieve sub-nanosecond accuracy. Even a clock drift of 1 nanosecond can result in a coordinate error exceeding 30 centimeters (assuming signal propagation at the speed of light in the environment)\cite{ref6}. 

\subsubsection{TDOA}

\begin{figure}[!t]
        \centering
        \includegraphics[width=0.7\linewidth]{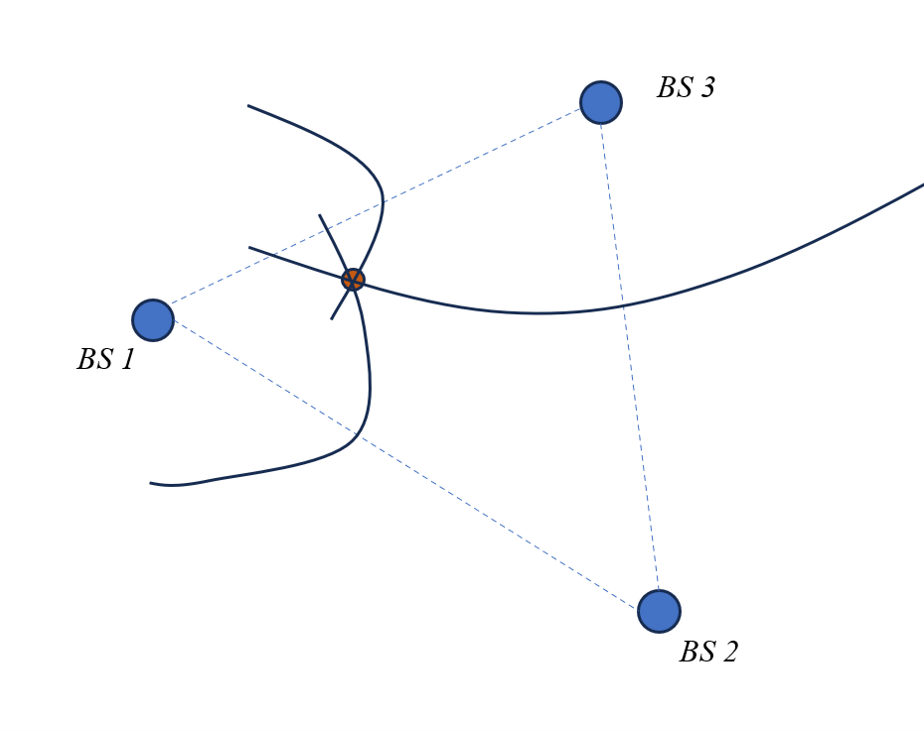}
        \caption{TDOA based positioning.}
        \label{fig2}
\end{figure}

The TODA algorithm reduces the requirements for clock synchronization by only necessitating time synchronization among base stations, without any specific requirements for the target. The working principle of it is to calculate the time difference for the arrival of signals from base stations to the target. And using this time difference to determine the difference in distances. As Fig.\ref{fig2} shows, by considering any two base stations, a set of possible trajectories (hyperbolic curves) where the target may be located can be derived. Furthermore, by identifying the intersection points of these hyperbolic curves, the position of the target can be estimated\cite{ref4}.

\subsubsection{AOA} 

The AOA algorithm is based on methodologies that leverage antenna arrays integrated into the receiver to ascertain the angle at which the transmitted signal intersects with the receiver. This estimation is accomplished by exploiting and computing the time difference of arrival at individual elements of the antenna array. One notable advantage of AOA is its capability to estimate the location of a device or user using a minimal number of monitors, namely two in a 2-D environment or three in a 3-D environment, respectively. Despite its potential for accurate estimation when the transmitter-receiver distance is short, AoA necessitates intricate hardware and meticulous calibration in comparison to Received Signal Strength (RSS) techniques\cite{ref111}. Additionally, its accuracy diminishes as the distance between the transmitter and receiver increases, whereby even a slight error in the angle of arrival calculation results in a substantial deviation in the actual location estimation.

\begin{equation}
        \left\{
                \begin{aligned}
                \tan \theta _1   &   =  & \frac{y-y_1}{x-x1}\\
                \tan \theta _2   &   =  & \frac{y-y_2}{x-x2}
                \end{aligned} \right. 
\end{equation}
Assuming the positions of the two base stations are $(x_1,y_1), (x_2,y_2)$, respectively. $\theta_1$, $\theta_2$ are the Angle at which the tag signal arrives at the base station. If there is a geometric relationship, the following equations can be obtained.

\section{From UWB to Terahertz Technology}
\label{section8}
The UWB technology can be traced back to the 1960 and the research on UWB technology has gradually matured. With an extremely wide spectrum characteristic, UWB communication allows data transmission across a wide range of frequency bands. Consequently, it could realize an extremely high transmission rate. However, since the frequency of UWB communication is limited to the range of 3.1GHz to 10.6GHz, and the bandwidth is also limited to the order of gigahertz, the transmission rate and development of UWB communication are subject to certain limitations.

Currently, the fifth-generation (5G) wireless network has been deployed in certain regions around the world, and it is expected to be fully used worldwide after a few years. However, although the 5G communication systems have offered significant improvements over the existing systems, they will not be able to fulfill the demands of future emerging intelligent and automation systems after ten years \cite{g1,g17}. Additionally, the capacity of 5G is anticipated to reach its limit by 2030 \cite{g2,gw2}. As a result, the terahertz frequency band, which possesses wide bandwidth resources, has become a new research hotspot in global communication research and is recognized as the key component of future 6G wireless communication systems \cite{g3,gw1}. Ultra-wideband terahertz communication (UWB-THz communication) is possible to realize the extremely high transmission rate by providing extremely wide bandwidth, reaching several hundred Gbps or even few Tbps. So this technology can meet the data rate requirements of future wireless local area network (WLAN) and wireless personal area network (WPAN) systems \cite{g4,g15}, having great potential and development value.

\subsection{Characteristics of Ultra-wideband Terahertz}
\begin{figure}[!t]
        \centering
        \includegraphics[width=0.99\linewidth]{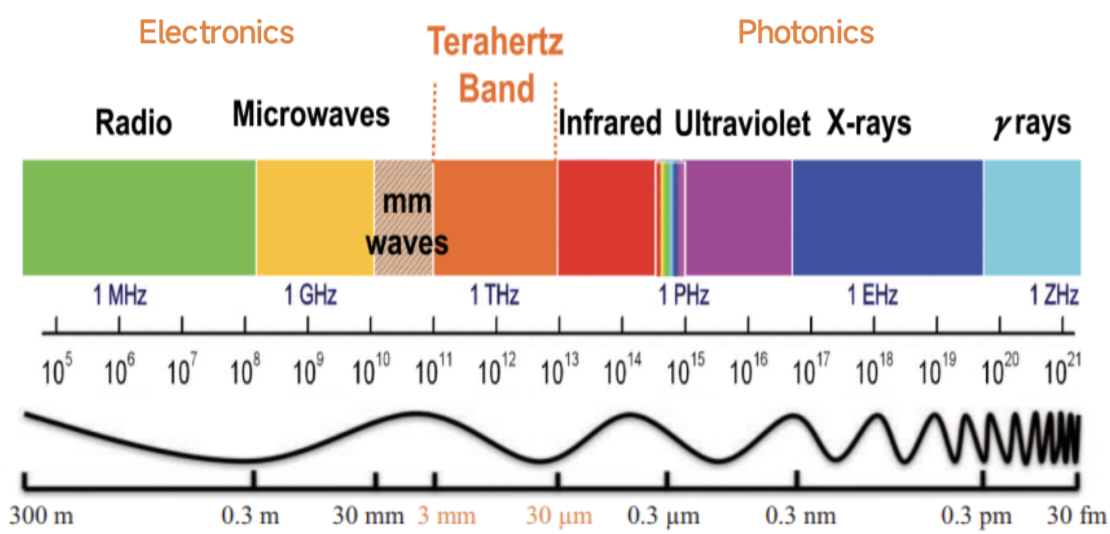}
        \caption{Electromagnetic spectrum\cite{g7}.}
        \label{g1}
        \end{figure}

As shown in Fig.\ref{g1}, the terahertz band is situated between two well-developed spectra: the electronic spectrum and the photonic spectrum. And its frequency is from 0.1 THz to 10 THz, while its wavelength is from 30 micrometers to 3 millimeters. Due to the lack of efficient terahertz sources and detection techniques in the past, 
and people were unable to produce efficient transceivers and 
antennas at terahertz frequencies, there is limited knowledge and 
use of this frequency band. Therefore the band used to be called “THz Gap". However, in the past decade, the rapid development of laser technology has provided stable and reliable excitation sources for generating terahertz waves. Additionally, research on terahertz wave detection techniques and their applications has been vigorously developed \cite{g6}, leading to continuous advancements in the research and application of the terahertz band. 

With an extremely wide bandwidth range, the frequency bandwidth of UWB-THz communication can cover a range from several GHz to several THz, which is much wider than traditional UWB communication. So UWB-THz communication can support data transmission rates of several hundred Gbps or even several Tbps, which is much faster than that of microwave communication and traditional UWB communication. And it will have strong anti-interference ability.

Considering traditional IR-UWB communication has the characteristics of strong anti-interference ability, strong confidentiality ability, high multipath resolution, and high transmission rates for its pulse width is very short, while the time of Terahertz pulse can reach the picosecond lever, which is shorter than the traditional UWB pulse, so the IR-UWB-THz communication has stronger anti-interference ability, stronger confidentiality ability, higher time resolution and higer multipath resolution.

In addition to the advantages of UWB communication, UWB-THZ communication also possesses some unique characteristics due to the unique characteristics of the terahertz band.

\begin{itemize}
\item Due to the wide frequency range of the terahertz band, and the fact that a large portion of it remains unallocated, UWB-THz communication benefits from abundant band resources. And the use of the terahertz band for communication can also effectively alleviate the increasingly tense spectrum resources and the capacity constraints of current wireless systems.

\item Because of the higher frequency and shorter wavelength of terahertz, UWB-THz communication have better directionality and are less susceptible to free-space diffraction than microwave and millimeter wave.

\item Due to the shorter wavelength of terahertz waves, antennas can be made in small, and other system structures can be made simpler and more economical while achieving the same functionality.

\item  Terahertz photon energy is lower than that of optical communication, about 1/40th of the photon energy, therefore, UWB-THz communication is more energy efficient than optical communication.

\item Due to the free space loss and molecular absorption phenomenon in the wireless space transmission process, UWB-THz communication can only be received at close range, making them more secure.

\item Terahertz waves have good penetrability to many dielectric materials and non-polar substances, and UWB-THz communication has a good ability to penetrate sand, dust and smoke, so it can carry out normal communication work in harsh environments with heavy dust and smoke.

\end{itemize}
\subsection{Challenges and Solutions}
\subsubsection{Challenges}
Despite the abundant band resources and high transmission rates, UWB-THz communication faces several challenges due to the limitations of the terahertz band:

Free space loss: Compared to the microwave band, the terahertz band has a higher frequency and shorter wavelength, leading to greater free space loss during wireless transmission, which not only restricts the transmission distance of terahertz communication, but also reduces the signal-noise ratio (SNR) of the system, thus affecting data capacity.

Molecular absorption: As most of the polar molecules' and organic molecules vibration and rotation energy level spacing are located in the terahertz band, especially water vapor in the air, in the terahertz band will show strong absorption and resonance phenomenon, the so-called molecular absorption, which will not only degrade the received power, but it can also intensify the noise\cite{g9}, greatly limiting the transmission distance of UWB-THz communication.

\subsubsection{Solutions}
Due to the prominent effects of free space loss and molecular absorption in the wireless space transmission process, the communication distance is limited. As a result, researchers in various countries have tried to take several measures to address this challenge. 

High gain antenna: The problem of larger path loss in the terahertz band can be solved by effective antenna technology and power control. Antenna technology is an important support in wireless communication to improve signal quality, reduce interference, generate diversity and reduce transmit power, which can concentrate the transmit power in the desired direction and can enhance the received signal-to-noise ratio of the receiver. Although higher transmission power can be achieved through power amplifiers, terahertz power amplifiers generally require significant direct current power consumption. Therefore, from the perspectives of energy efficiency and performance enhancement, adopting efficient antenna technology is more suitable for terahertz wireless communication systems. And due to the extremely high frequency and extremely short wavelength of terahertz waves, it is possible to deploy multiple antennas in the same physical area. Moreover, the deployment of high gain antennas in terahertz communication can further enhance the ability to combat high path loss\cite{g10}.
\begin{figure}[!t]
        \centering
        \includegraphics[width=0.9\linewidth]{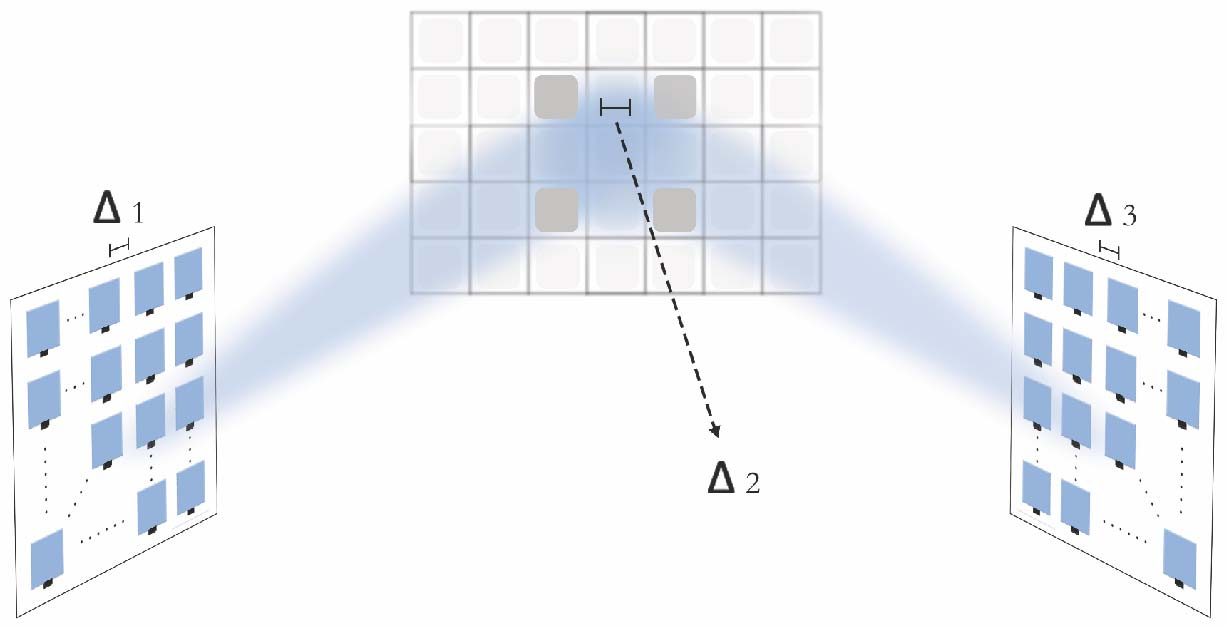}
        \caption{THz communications assisted by IRS\cite{g12}.}
        \label{g2}
\end{figure}

Intelligent reflecting surfaces(IRS): As shown in Fig.\ref{g2}, an IRS consisting of a large number of reconfigurable passive elements can be applied to change the propagation direction and enhance the signal strength by adjusting the phase or amplitude of the IRS elements, thereby improving the coverage of the THz signals\cite{g11}.

\begin{figure}[!t]
        \centering
        \includegraphics[width=0.99\linewidth]{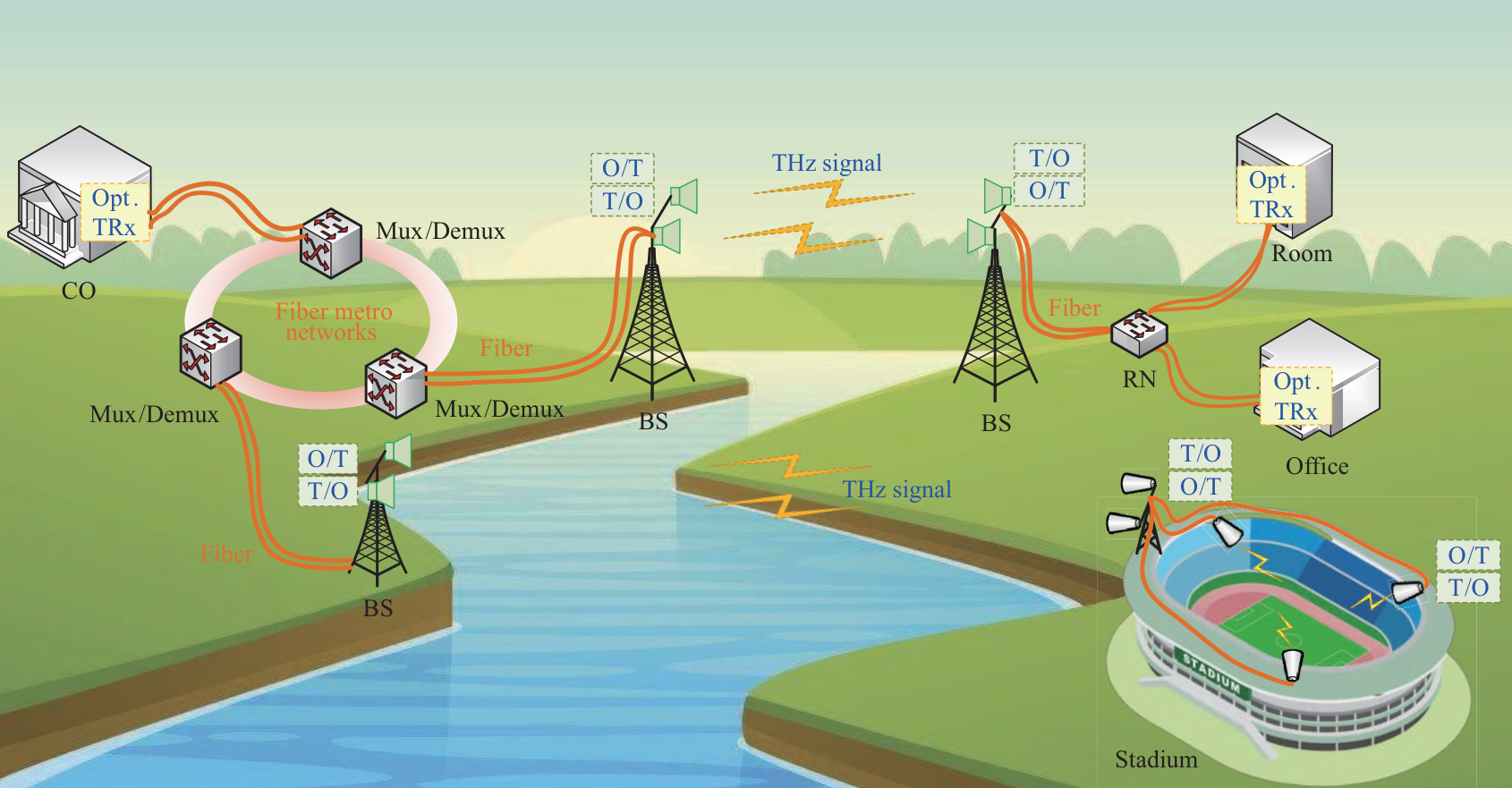}
        \caption{Fiber-THz-Fiber seamlessly communication system\cite{g14}.}
\label{g3}
\end{figure}

Fiber-THz-fiber communication system: In view of the inflexibility and high price of the optical fiber communication system, UWB-THz communication can become an attractive complementary technology to the optical fiber connections\cite{g5,g13}. As shown in Fig.\ref{g3}, the ultrahigh-speed fiber-THz-fiber communication system based on the photonic upconversion and hybrid optoelectronic down-conversion techniques can enable the collaborative work between terahertz and fiber optic communications \cite{g14}, which can not only expand the coverage and application scenarios of UWB-THz communication, but also enhance the flexibility of optical fiber communication.
\section{Conclusion}
UWB technology is a promising wireless communication solution that boasts distinctive features and vast potential for diverse applications. By generating and transmitting radio-frequency energy across an extensive frequency range, UWB achieves high data rates and precise ranging and positioning capabilities, making it suitable for indoor positioning, communication systems, that’s sort of things. 

Looking ahead, the demand for greater frequency bandwidth is undeniably set to increase as we venture into higher and broader frequency bands to support faster data transmission and accommodate a wide array of applications. Alongside the progress in wireless communication technology and the relentless pursuit of improved communication performance, the integration of UWB technology has also given rise to Terahertz communication in various fields. However, this upward trend presents its own set of obstacles, such as signal attenuation and reduced penetration capabilities in higher frequency ranges, as well as challenges associated with limited coverage areas. Overcoming these hurdles will necessitate relentless technological innovation.

Overall, UWB, as an exemplary communication technology, offers a plethora of advantages with its wide bandwidth, and its continuous evolution through synergizing with other technologies promises to unlock new value in the future.

\bibliographystyle{IEEEtran}
\bibliography{references}

\end{document}